\pdfoutput=1
\documentclass[aps,prl,floatfix,twocolumn]{revtex4}
\usepackage{amsfonts}
\usepackage{amsmath}
\usepackage{graphicx}
\usepackage{color}
\usepackage{epstopdf}
\usepackage{graphicx}
\usepackage{latexsym}
\usepackage{units}

\begin{document}
\title{The secret world of shrimps: polarisation vision at its best}
\author{Sonja Kleinlogel$^{1,2}$ and Andrew G. White$^{3}$}
\affiliation{$^{1}$ Max Planck Institut f\"{u}r Biophysik, Frankfurt, 60438 Germany\\
$^{2}$ School of Biomedical Sciences, University of Queensland, Brisbane, QLD 4072, Australia\\
$^{3}$ Department of Physics and Centre for Quantum Computer Technology, University of Queensland, Brisbane, QLD 4072, Australia \vspace{0 mm}}

\begin{abstract}
\noindent Animal vision spans a great range of complexity, with systems evolving to detect variations in optical intensity, distribution, colour, and polarisation. Polarisation vision systems studied to date detect one to four channels of linear polarisation, combining them in opponent pairs to provide intensity-independent operation. Circular polarisation vision has never been seen, and is widely believed to play no part in animal vision. Polarisation is fully measured via Stokes' parameters---obtained by combined linear and circular polarisation measurements. Optimal polarisation vision is the ability to see Stokes' parameters: here we show that the crustacean \emph{Gonodactylus smithii} measures the exact components required.  This vision provides optimal contrast-enhancement, and precise determination of polarisation with no confusion-states or neutral-points---significant advantages. We emphasise that linear and circular polarisation vision are not different modalities---both are necessary for optimal polarisation vision, regardless of the presence of strongly linear or circularly polarised features in the animal's environment.
\end{abstract}
\maketitle

Polarisation is the plane of vibration of the electric field vector of light. In unpolarised light, the plane changes completely randomly with time, if the plane changes predictably with time the light is said to be fully polarised. In nature the ambient---scattered and/or reflected---light tends to be partially-polarised, lying between these extremes. Fully polarised light ranges from \emph{linear}, where the plane is constant with time, through elliptical to \emph{circular}, where the plane rotates 360$^\circ$ every optical period (with respect to the propagation axis). Optimal polarisation vision is the ability to measure \emph{all} aspects of polarisation in the visual field. In optics, the state of polarisation is plotted by a vector resting on the surface of, or in, a sphere called the Poincar\'e sphere, Figure 1. Vectors that rest on the surface of the sphere represent fully polarised light, shorter vectors represent partially-polarised light and the centre of the sphere represents unpolarised light. In rectangular coordinates, the vector position is given directly by Stokes' parameters \cite{Stokes, Hecht, BW}, for example,
\begin{equation}
S_{1} {=}   \frac{I_{h} {-} I_{v}}{I_{h} {+} I_{v}}, \ \ S_{2} {=} \frac{I_{d} {-} I_{a}}{I_{d} {+} I_{a}}, \ \ S_{3} {=}  \frac{I_{r} {-} I_{l}}{I_{r} {+} I_{l}}
\end{equation}
where $I$ is intensity; $\{ h,v,d,a \}$ represent horizontal, vertical, diagonal and anti-diagonal linearly polarised light; $\{ r,l \}$ represent right- and left- hand circularly polarised light; and Stokes' parameters are normalised to unity for convenience. (The Stokes' parameter for total light intensity, $S_{0}$, contains no information about the polarisation state and so we do not consider it here). A common alternative is to describe the Stokes' vector in spherical coordinates: its length is the \emph{degree} of polarisation, $\mathcal{P}$, the angles $\theta$ and $\varphi$ indicate the \emph{type} of polarisation. In biological parlance, each of the above six polarisation components is a separate channel: optimal polarisation vision requires measurement of all three Stokes' parameters, i.e. all six polarisation channels. Optimal polarisation vision confers obvious advantages to the possessor: detection of any change in the degree and type of polarisation---without needing assumptions about the polarisation background---even if the object causing that change is effectively invisible without the polarisation information.

Polarised light is abundant in nature. Visual backgrounds can be partially-polarised by scattering of natural light in the atmosphere or under water, or by reflection from natural surfaces such as the shiny cuticles of leaves or the air/water interface \cite{Ivanoff58,Horvath03,Wehner01,Cronin01}. Background light can be polarised by biological surfaces, for example reflection from birefringent arthropod cuticles \cite{Neville71, Horvath06} or  scattering from marine phytoplankton \cite{Shapiro91}; and by transmission, for example through the semi-transparent bodies of dinoflagellates \cite{Shapiro91}. Biological entities can also emit polarised light, for example fluorescent light emitted from chlorophyll \cite{Gafni75}, or the left and right lanterns of firefly larvae which emit left- and right- circularly polarised light \cite{Wynberg80}. 

Linear polarisation sensitivity in arthropods and its biological implications have been studied intensely since the 1950s. A well known example is the use of the skylight polarisation pattern by arthropods for navigation and orientation \cite{Wehner81}. Sensitivity to a single linear polarisation component increases contrast \cite{Wehner01,Schechner04}; sensitivity to two or more linear polarisation components has been implicated in a range of visual functions including orientation \cite{Waterman88}, navigation \cite{Goddard91, Schwind99}, prey detection \cite{Shashar98, Sabbah05}, predator avoidance \cite{Shashar00} and intra-species signaling \cite{Cronin03}. Biologists are aware that circularly-polarised light is rare in nature, and a common conclusion is, to quote from the standard text on polarised light in animal vision \cite{Horvath03}: ``Thus, it is questionable whether circular/elliptical polarisation of light in nature could have any biological importance''. Since optimal polarisation vision \emph{requires} simultaneous measurement of linear \emph{and} circular polarisation, as discussed above, there is a clear advantage in evolving the ability to detect both.

All crustaceans have the ability to sense linear polarisation over the whole compound eye, which is composed of several hundred visual units, the \emph{ommatidia}. Each ommatidium consists of a cornea covering a lens, behind which lie eight photoreceptors, called \emph{retinular cells}, clustered around a light guide, the \emph{rhabdom}, see Fig \ref{rhabdom}a. The first cell,  centrally positioned around the light guide, is a small ultraviolet-sensitive retinular cell \cite{Marshall99b}, R8. Under this are seven retinular cells, sensitive in the visible, that run the length of the light guide, R1--7, see Figure \ref{rhabdom}. The retinular cells extend parallel microvilli into the light guide: in R1--7 the microvilli alternate in orthogonal layers down the entire length of the light guide, between a group of three cells (group I = R1, R4, R5) and a group of four cells (group II = R2, R3, R6, R7). The microvilli  contain rhodopsin---a pigment molecule with a strong dipole moment---and are narrow tubes, ${\sim}60$ nm in diameter, Fig.~\ref{eye}c, aligning the rhodopsin so that the retinular cells act as linear polarisation sensors \cite{Horvath03,Kleinlogel06}.

One group of crustaceans, the stomatopods, have evolved an equatorial mid-band in their eyes, see Figure \ref{eye}a,b. The resulting dorsal and ventral hemispheres (DH and VH) each sense linear polarisation, but rotated 45$^{\circ}$ with respect to each other; the mid-band, a section of between 2 and 6 rows of ommatidia, is specialised for colour, or polarisation, or both \cite{Marshall88}. In particular, \emph{Gonodactylus smithii} and other gonodactyloid stomatopod species possess six-rowed mid-bands where the first four rows are specialised colour receptors (11 visual pigments \cite{Marshall88,Marshall99b}, spanning 290 to 730 nm). In this paper we show by means of intracellular recordings that the two mid-band rows 5 and 6 are specialised for detecting \emph{circular} polarisation, the first reported circular polarisation sensitivity in any organism. We further show that stomatopods have both the anatomical and the neuronal features to measure all StokesÕ parameters---essential for optimal polarisation vision.

The structure of the rhabdoms in rows 5 and 6 are similar to that in the dorsal and ventral hemispheres with a few key differences: the R1-7 rhabdom is wider and more crystalline in appearance; the microvilli layers are thinner; and the ultraviolet sensitive cell on the top, R8, is structurally unusual. It is a four-lobed cell that surrounds the light guide; it's ovoid in transverse section and extends substantially further along the light guide then in the rest of the eye; and it is anisotropic, i.e. there is a preferred direction set by parallel microvilli extending between the lobes of the cell as shown in Figure 2D. The optical axis of the R8 cell (indicated by the microvilli orientation) is oriented at 45$^{\circ}$ to the detection axes of cells R1--7. Fig.~\ref{rhabdom}d shows the arrangement for row 5; row 6 maintains the same relative arrangement but the entire rhabdom is rotated by 90$^{\circ}$ (counter-clockwise when seen from front in right eye). It is these structural properties of the R8 cell that introduces a relative phase shift to orthogonal polarisations that pass through it, As we will show later, the R8 cells of rows 5 and 6 almost perfectly convert circularly polarised light to linearly polarised light, which is then detected by the alternating stacks of microvilli produced by R1--7. We hypothesise that the R8 cells have evolved to secondarily act as quarter-wave retarders in the visible, ${\sim}400{-}700$ nm.

Of course, anatomical structure can only indicate possible function: we tested function directly using intracellular electrophysiology. The eye was mounted so that the lateral mid-band was aligned approximately to the horizontal. A sharp electrode was inserted through a hole cut into the dorsal cornea, then impaled into the photoreceptor under test. The receptors were illuminated with \unit[50]{ms} flashes of light from a Xenon arc lamp passed through a UV cut-off filter, giving a test spectrum of ${\sim}400{-}700$ nm. The spectral sensitivities were measured with unpolarised light: we used a spectral scan method where a photoreceptor is clamped to a pre-selected DC potential by adjusting the light flux as we scanned from 300 to 730 nm in 10 nm steps with a monochromator. Fig. \ref{data} shows the average spectral sensitivity for: \emph{top} dorsal and ventral hemispheric photoreceptors; and \emph{bottom} mid-band rows 5 and 6 photoreceptors. Note that the response for both is broad and very similar: the cells are near homochromatic in the visible. In both cases the photoreceptor response declines steeply above 600 nm; the significant difference between the spectra is the UV peak for the hemispheric cells. This may be due to: electrical coupling of R8 to the R1--7 cells in the hemispheric case \cite{Kleinlogel05}; suppressed UV response in the mid-band case to absorption by the extra-ordinarily long mid-band R8 cells (22\% of rhabdom!); or a combination of the two. 

The light was polarised with a combination of a linear polariser and a broadband quarter-wave plate, effective from 450--610 nm. Figure \ref{data}b shows the temporal response of the photoreceptors to the 50 ms flash of light, in this case for left- and right- circularly polarised light. The time curves are used to determine the peak responses, used in the remaining data analysis. Figures \ref{data}b,c show the raw angle-response data for the R1 photoreceptors (group I), respectively of the dorsal and ventral hemispheres (the dye-injected cell shown in Fig. \ref{rhabdom}c). All angles are given relative to the 0$^\circ$ (vertical) position of the linear polarisation filter. The linear polarisation sensitivity was determined by stimulation with flashes of polarised light varied in angular steps of 10$^\circ$. 

It is clear that the R1 receptors respond strongly to linearly polarised light: the minimum responses are non-zero since the photoreceptors are not perfect polarisation sensors, as we show below. Note that the dorsal receptor has a maximum response at 60$^\circ$; the ventral at 105$^\circ$---reflecting the 45$^\circ$ difference in microvilli orientation. This of course is true for all group I receptors; for group II receptors the response will be moved by 90$^\circ$. (See online material Figure 2).

Fig \ref{eye}b shows the resting position of the eye. The three dark areas are the so-called pseudo-pupils, indicating that light from the direction of the viewer is being strongly absorbed in those areas---that is, the three pseudo-pupils simultaneously share the visual field. The sharp electrode was inserted in approximately this region of the eye where the rows of ommatidia are inclined ${\sim}10{-}20^\circ$. The results for Fig. \ref{data}c,d indicate that cells were recorded from ommatidia inclined at 15$^{\circ}$ to the horizontal. For convenience hereafter we will refer to 105$^{\circ}$ and 15$^{\circ}$ polarised light as horizontal and vertical $(h, v)$ and 150$^{\circ}$ and 60$^{\circ}$ polarised light as diagonal and anti-diagonal $(d, a)$, respectively.

Photoreceptor responses are intensity-dependent, with a logarithmic response saturating at higher light intensities, the curves in Fig. \ref{data}c,d are fitted logarithmic square cosines, see caption for details. We quantify the saturation by taking intensity-response data, Fig. \ref{data}e,f, which plots the response (mV) vs the relative light intensity, $\log(I/I_{0})$. There are two lines of data: the upper are taken at the polariser angle corresponding to maximum response, $\phi_{\mathrm{max}}$, the lower at the angle for minimum, $\phi_{\mathrm{min}}$. There is a linear response region centred at the half-maximum: the difference in relative intensity on a logarithmic scale, $\delta_{i}$, gives the polarisation sensitivity \cite{Kleinlogel06}, $10^{\delta_{i}}$. The measured polarisation sensitivities are large, 9.44$\pm$0.02 and 10.56$\pm$0.02, respectively for the dorsal and ventral R1 cells. These sensitivities are comparable to the high values of 7-12 measured in crabs \cite{Shaw66} and crayfish \cite{Waterman70}.

Circular polarisation sensitivity was determined by stimulating photoreceptors in rows 5 and 6 with flashes of left- and right- circularly polarised light. Thus for example, Figure \ref{data}b shows the response for cell R3 in row 5, the stained cell in Fig. \ref{rhabdom}d. The cell clearly responds more strongly to left- than right- circularly polarised light. Polarisation sensitivity was once again determined from intensity-response data, Fig. \ref{data}g shows the circular polarisation sensitivity for a row 5 R1 cell, $10.84{\pm}0.02$, comparable with the linear polarisation sensitivities measured above. It is possible that this cell is sensitive in some degree to linearly polarised light, to check this we sent in diagonal and anti-diagonal linear polarisation, aligned with the microvilli of the R1--7 cells. Figure \ref{data}h shows the result---to within error there is zero linear polarisation sensitivity. This suggests that the R8 cell acts effectively as a quarter-wave retarder across the test spectrum.

We can determine the exact polarisation state that each cell is sensitive to using polarisation tomography \cite{James01}, i.e. sending in the set of states $\{h,  v,  d,  a, r,  l\}$, measuring the response for each, and using these to calculate Stokes' parameters. Table \ref{Stokes} shows the results for R1 cells measured in the dorsal and ventral hemispheres, and row 5 from the mid-band, Figure \ref{data}c-g. Respectively, each cell most strongly responds to diagonal, horizontal and right-circular polarised light, as evidenced by the dominant $S_{2}$, $S_{1}$, and $S_{3}$ parameters. It is also clear that the cells are acting as partial-polarising detectors: this is quantified by the degree of polarisation, $\mathcal{P}$, which is 1 for a perfect polariser. Averaging over ten hemispheric retinular cells (dorsal and ventral, groups I \& II) we find an average value of $\overline{\mathcal{P}}_{\mathrm{hemi}}{=}0.145{\pm}0.035$; for nine mid-band cells (rows 5 and 6, groups I \& II) we find $\overline{\mathcal{P}}_{\mathrm{mid}}{=}0.340{\pm}0.061$ (full data in online material Table I). We see that the mid-band cells give a much larger signal for totally polarised light then the hemispheric cells; this is consistent with the observations that the microvilli in the mid-band are more ordered and in thinner layers, Fig.~\ref{eye}c, which is expected to reduce self-screening and give a better polarising response. Regardless of the strength of the response, a crucial ability is to preferentially distinguish just one of Stokes' axes: this is measured by the discrimination, $\mathcal{D}{=}(S_{i}/\mathcal{P})^{2}$, where $i{\in}\{1,2,3\}$. Our measurements shows that the photoreceptor cells have near-perfect discrimination, Table I.

\emph{Gonodactylus smithii} thus has all the requirements for optimal polarisation vision. Each eye possesses four linear ($h,  v,  d,  a$) and two circular ($r,  l$) polarisation input channels, which are homochromatic, Fig. \ref{data}a, and acquire data simultaneously, since they share the same visual field, Fig. \ref{eye}b. There exists striking structural \cite{Kleinlogel05} and behavioural \cite{Marshall99b} evidence for opponent circuitry between the orthogonal polarisation channels within the eyestalk. That is, the neural signal from one channel is subtracted from the other. This is essential for Stokes' parameter analysis. Polarisation vision in stomatopods has mainly been implicated with intra-specific signal recognition, since many  species reflect polarised light from their bodies \cite{Cronin03}. However, the carapace of \emph{Gonodactylus smithii} does \emph{not} reflect linear or circular polarised light---polarisation vision in this species is clearly being used for something else. Stomatopods are shallow-water crustaceans in a visual environment with a partially-polarised background \cite{Ivanoff58,Cronin01}. Crustaceans are known to use polarisation for navigation; many stomatopod prey species are either reflective or transparent but change the polarisation of the light \cite{Sabbah05,Shashar98,Shashar00}---an obvious possible driver of evolutionary change. Optimal polarisation vision provides \emph{all} the information about polarisation of the visual field---giving the greatest ability to detect changes in both the \emph{degree} and \emph{type} of polarisation. This goes beyond simple contrast enhancement: optimal polarisation vision is analogous to the improvement afforded by stereo over mono vision in terms of increased information capacity.

Humanity began to use polarisation vision only recently---perhaps dating back to Viking use of 
Icelandic feldspar to navigate on cloudy days \cite{Hegedus07}---our move to optimal polarisation vision is significantly more recent, requiring three-channel camera systems and fast computer software. Once again, Nature seems to have anticipated our best technological advances, with \emph{Gonodactylus smithii} being the first organism described with the physiological and neurological components necessary for optimal polarisation vision. We feel it is worth re-examining other organisms for similar visual systems, and widening the use of machine-based optimal polarisation vision systems in both field and laboratory biology---so we too, can begin to see the world as shrimps do.

\vspace{5 mm}
\small
We acknowledge valuable discussions with Thomas Labhart, Alexei Gilchrist and Michael Harvey. Justin Marshall and David Vaney provided help with the electrophysiological apparatus.  Financial support was provided in part by the Australian Research Council Discovery Project and Federation Fellow programs, the American Air Force (AOARD/AFOSR) (FA5209--04--P--0395), and the Swiss National Science Foundation (PBSKB--104268/1). Please address correspondence to SK (sonja.kleinlogel@mpibp-frankfurt.mpg.de). 

\begin{table}
\begin{center}
\begin{tabular}{|c|c|c|c|} \hline
			 		& R1, Dorsal			& R1, Ventral 		& R1, Row 5 \\ \hline
$S_{1}$		& $\phantom{-}0.015{\pm}0.012 \ $	& $\phantom{-}0.196{\pm}0.015 \ $
												& ${-}0.024{\pm}0.011 \ $  \\ 
$S_{2}$		& ${-}0.189{\pm}0.014$			& $\phantom{-}0.012{\pm}0.014 \ $
												&  $\phantom{-}0.015{\pm}0.011 \ $  \\ 
$S_{3}$		& $\phantom{-}0.000{\pm}0.012 \ $	& $\phantom{-}0.012{\pm}0.014 \ $
												&  $\phantom{-}0.434{\pm}0.016 \ $  \\ \hline
$\mathcal{P}$	& $\phantom{-}0.190{\pm}0.014 \ $ 	& $\phantom{-}0.196{\pm}0.026 \ $
												&  $\phantom{-}0.436{\pm}0.016 \ $ \\
$\theta$	& $-85.5{\pm}3.6^{\circ} \ $ 	& $\phantom{-}3.5{\pm}4.1^{\circ} \ $
												&  $-31{\pm}22^{\circ} \ $ \\
$\varphi$	& $\phantom{-}0.0{\pm}3.6^{\circ} \ $ 	& $\phantom{-}3.6{\pm}4.0^{\circ} \ $
												&  $\phantom{-}86.2{\pm}1.5^{\circ} \ $ \\\hline
$\ \mathcal{D}_{h,v} \ $ & $\phantom{-}0.006{\pm}0.010 \ $ & $\phantom{-}0.992{\pm}0.012 \ $
												& $\phantom{-}0.003{\pm}0.003 \ $ \\
$\ \mathcal{D}_{d,a} \ $ & $\phantom{-}0.994{\pm}0.010 \ $ & $\phantom{-}0.004{\pm}0.009 \ $
												& $\phantom{-}0.001{\pm}0.002 \ $ \\
$\ \mathcal{D}_{r,l} \ $   & $\phantom{-}0.000{\pm}0.000 \ $ & $\phantom{-}0.004{\pm}0.009 \ $
												& $\phantom{-}0.996{\pm}0.003 \ $ \\ \hline
\end{tabular}
\end{center}
\vspace{-3mm}
\caption{Table of Stokes' parameter responses for individual R1 cells measured in the dorsal and ventral hemispheres, and row 5 in the mid-band. The Stokes' parameters, $\{S_{1},S_{2},S_{3}\}$, are the rectangular coordinates of the Stokes' vector in the Poincar\'e sphere (unity radius for normalised Stokes' vectors). The length of the vector is the degree of polarisation, $\mathcal{P}{=}\sqrt{S_{1}^{2}{+}S_{2}^{2}{+}S_{3}^{2}}$, the spherical coordinates $\theta{=}\arctan(S_{2}/S_{1})$ and $\varphi{=}\arcsin(S_{3}/\mathcal{P})$ indicate the type of polarisation. For linearly polarised light, $\varphi{=}0^{\circ}$; for circularly polarised light. $\varphi{=}90^{\circ}$. The R1 cells act as partially-polarising detectors, with mid-band cells being better polarisers than hemispheric cells. The ability of each cell to distinguish along one of the Stokes' axes is given by the discrimination, $\mathcal{D}{=}(S_{i}/\mathcal{P})^{2}$, where $i{\in}\{1,2,3\}$. Dorsal photoreceptor cells respond most strongly to diagonal/anti-diagonal linear polarisation; ventral to horizontal/vertical; and mid-band to right/left circular polarisation.
\vspace{-7mm}}
\label{Stokes}
\end{table}

\newpage
\vspace{1 mm}
\noindent \textbf{Materials and Methods}. 

\noindent \textsf{Animals and preparation}. Adult male and female stomatopods of the species \emph{Gonodactylus smithii} (Crustacea, Hoplocarida, Stomatopoda, Gonodactyloidea) were collected with hand-nets from reef flats on Lizard Island (Queensland, Australia, GBRMPA permit \# G06/15528.1) and were maintained under a 12h:12h dark/light cycle in marine aquaria approved by AQIS (Australian Quarantine Inspection Service) and Environment Australia Wildlife Protection. Animals were anaesthetized by cooling before the eyes were removed and the animal euthanized by decapitation. All procedures were approved by the Animal Ethics Committee (UAEC, permit \# VTHRC/488/06) of the University of Queensland.

The amputated eye was mounted on a plastic rod with the lateral mid-band region oriented horizontally and immersed in oxygenated stomatopod saline (Fig. 1 supplement). The preparation was placed at the centre of a cardan arm arrangement carrying the end of a liquid light guide supplying a 0.9$^{\circ}$ light stimulus, produced by a 150 W Xenon-arc lamp (Oriel, Stratford, USA) in combination with a computer-controlled monochromator (Oriel, Stratford, USA). At the location of the eye the white light had an unattenuated maximal intensity of approximately $10^{18}$ photons s$^{-1}$ cm$^{-2}$, which could be adjusted with a computer-controlled neutral density wedge (0-4 on a relative logarithmic scale, Edmund Optics). 

\vspace{1mm}
\noindent \textsf{Electrophysiology}. Microelectrodes either filled with 1\% Ethidium bromide (approx. 95\% HPLC, Sigma-Aldrich, St. Louis, MO, USA) in 1 M KCl (40-100 M$\Omega$) or 5\% Lucifer Yellow CH (Sigma-Aldrich Pty Ltd, Castle Hill, NSW, Australia) in 0.1 M Tris buffer and 1M LiCl (100-250 M$\Omega$) were lowered vertically into the retina through a corneal hole cut with a razorblade in the lateral dorsal hemisphere (Fig. 1 supplement). The pipette was connected to the headstage of an intracellular amplifier (Axoprobe 1A, Axon Instruments Ltd, Inverurie, Scotland) via a chloride silver electrode and an Ag/AgCl pellet immersed in saline served as ground electrode. Single photoreceptor responses were digitized on a virtual oscilloscope (ADC-100) using Pico Scope software (Pico Technology, Camperdown, NSW, Australia) and then exported into Microsoft Excel for analysis.

After impalement of a photoreceptor and approximate alignment of the light source with its optical axis, the receptor was characterized by its spectral sensitivity, which was measured with the spectral scan method \cite{Menzel}. In order to determine the linear polarisation sensitivity of the cell a UV-transmitting linear polarisation filter (HNPÕB, Polaroid Company) was inserted between the liquid light guide and the eye and its angle relative to the eye changed in angular steps of 10$^{\circ}$ (0$^{\circ}$ is vertical polarisation) whilst the eye was stimulated with brief (50 msec) flashes of light at 5 sec intervals. Two intensity-response R-(log I) functions were then recorded by applying 0.25 log intensity series of light-flashes at the two polarizer angles which elicited maximal ($\phi_{max}$) and minimal ($\phi_{min}$) photoreceptor response respectively. In order to deliver sufficient light to the photoreceptor---yet eliminating responses from the potentially electrically coupled overlying R8 cells---we used white light in combination with a 400 nm long-pass filter (1$\frac{1}{4}$'' UV/IR-Cut-Filter, Baader Planetarium, Mammendorf, Germany, transmission 400--700 nm) for stimulation.

To assess the circular polarisation sensitivity of the cell, an achromatic quarter wave retarder plate characterized by a practically constant absorption spectrum for wavelengths from $450{<} \lambda {<}610$ nm (Edmund Optics, Singapore) was inserted between the linear polarisation filter and the eye. To produce right- and left-handed circularly polarized light, the optical axis of the wave plate was oriented at -45$^{\circ}$ or +45$^{\circ}$ relative to the optical axis of the linear polarisation filter. Two R-(log I) functions were then recorded using flashes of left-handed and right-handed circularly polarized light respectively. At the end of each recording, cells were iontophoretically marked with either Lucifer yellow CH using a 0.8 to 1 nA hyperpolarizing DC current at 1 Hz for 4 to 5.5 min or with Ethidium bromide using a 0.6 to 1 nA depolarising DC current at 1 Hz for 3.5 to 4 min. Data sets were only accepted if there was no appreciable change in $\phi_{max}$ or resting membrane voltage during the set of runs. Polarisation responses computed were always above threshold and below saturation. Responses to $\{ h, v, d, a, r, l\}$ for Stokes' parameters were always measured in the linear response part of the R-(log I) curves. Only approximately parallel R-(log I) curves were used for analysis, since the principle of univariance applies \cite{Laughlin}. 

\vspace{1mm}
\noindent \textsf{Histology}. Eyes were fixed in 4\% paraformaldehyde and embedded in 2-hydroxyethylmethacrylate (Technovit T7100, Heraeus, Germany). Serial frontal plastic sections of 7 $\mu$m thickness were viewed under a Zeiss Axioscope microscope (10$\times$/0.30 and 20$\times$/0.5 objectives) equipped with a digital SPOT camera (Diagnostic Instruments, Sterling Heights, MI, USA) using fluorescent microscopy and ALPHA Vivid standard Lucifer yellow XF14 filters (Omega Optical, Inc., Brattleboro, VT, USA). Images were processed and enhanced in contrast using Adobe Photoshop 7.0 (Adobe Systems).

\vspace{1mm}
\noindent \textsf{Terminology}. To simplify the description of the eyeÕs anatomy, in particular the direction of microvilli to the outside world and the directions of maximal linear polarisation sensitivities ($\phi_{max}$) of individual photoreceptors, the text and all figures describe the directions as seen in a frontal view of a right eye with the mid-band arranged horizontally. In a left eye, the photoreceptor arrangement, microvillar orientations and $\phi_{max}$ will be mirror-symmetric.

\begin{table}
\begin{center}
\includegraphics[width=\linewidth]{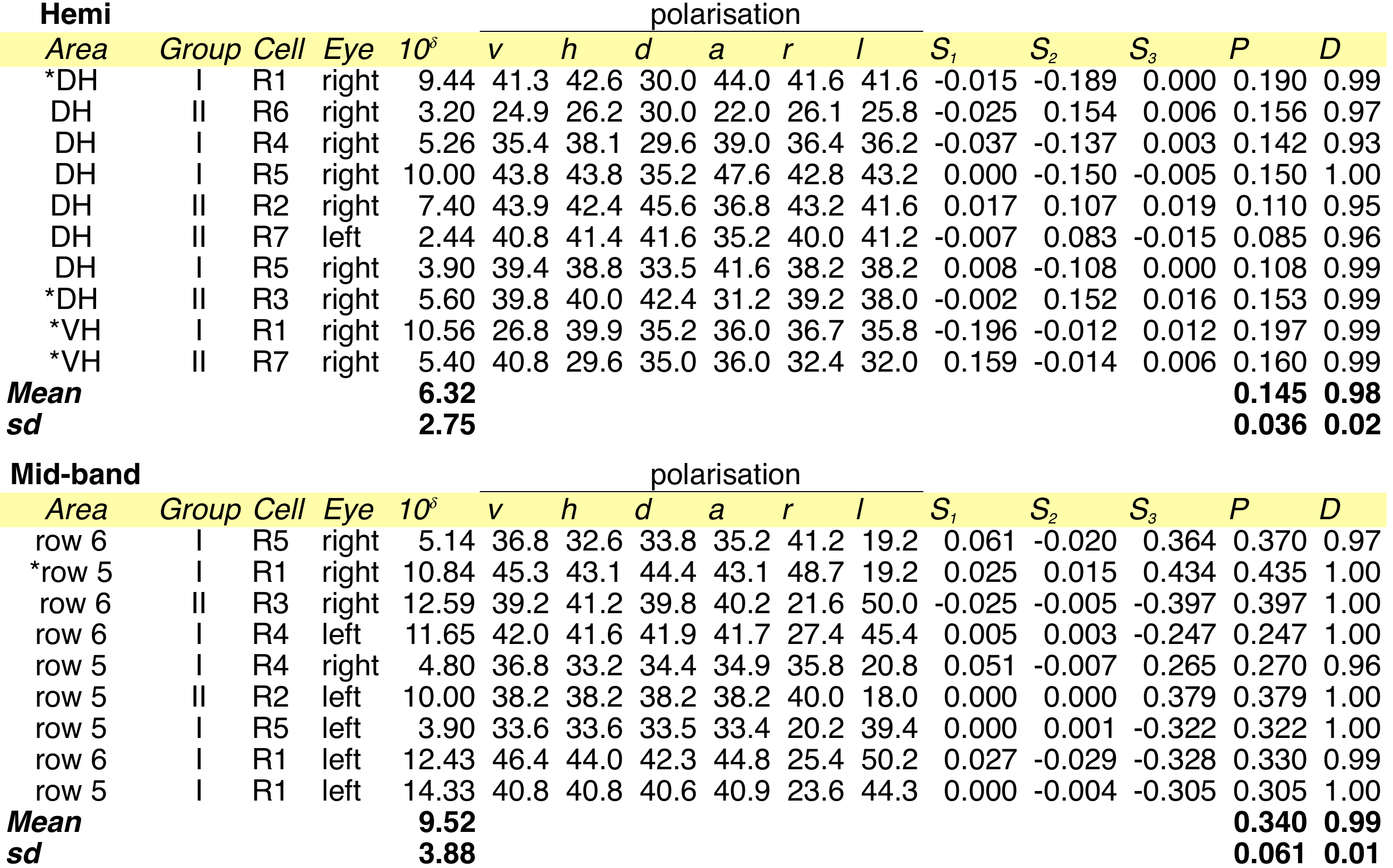}
\caption{Values of polarisation sensitivity, $10^{\delta}$, and peak responses and their derived quantities: Stokes' parameters, $S_{i}$, degree of polarisation, $\mathcal{P}$, and discrimination, $\mathcal{D}$. Entries indicated with * are shown in Figure 4 (main text) and Figure 2 (supplementary material).}
\label{supptable}
\end{center}
\end{table}

\begin{figure*}
\begin{center}
\includegraphics[width=0.5 \linewidth]{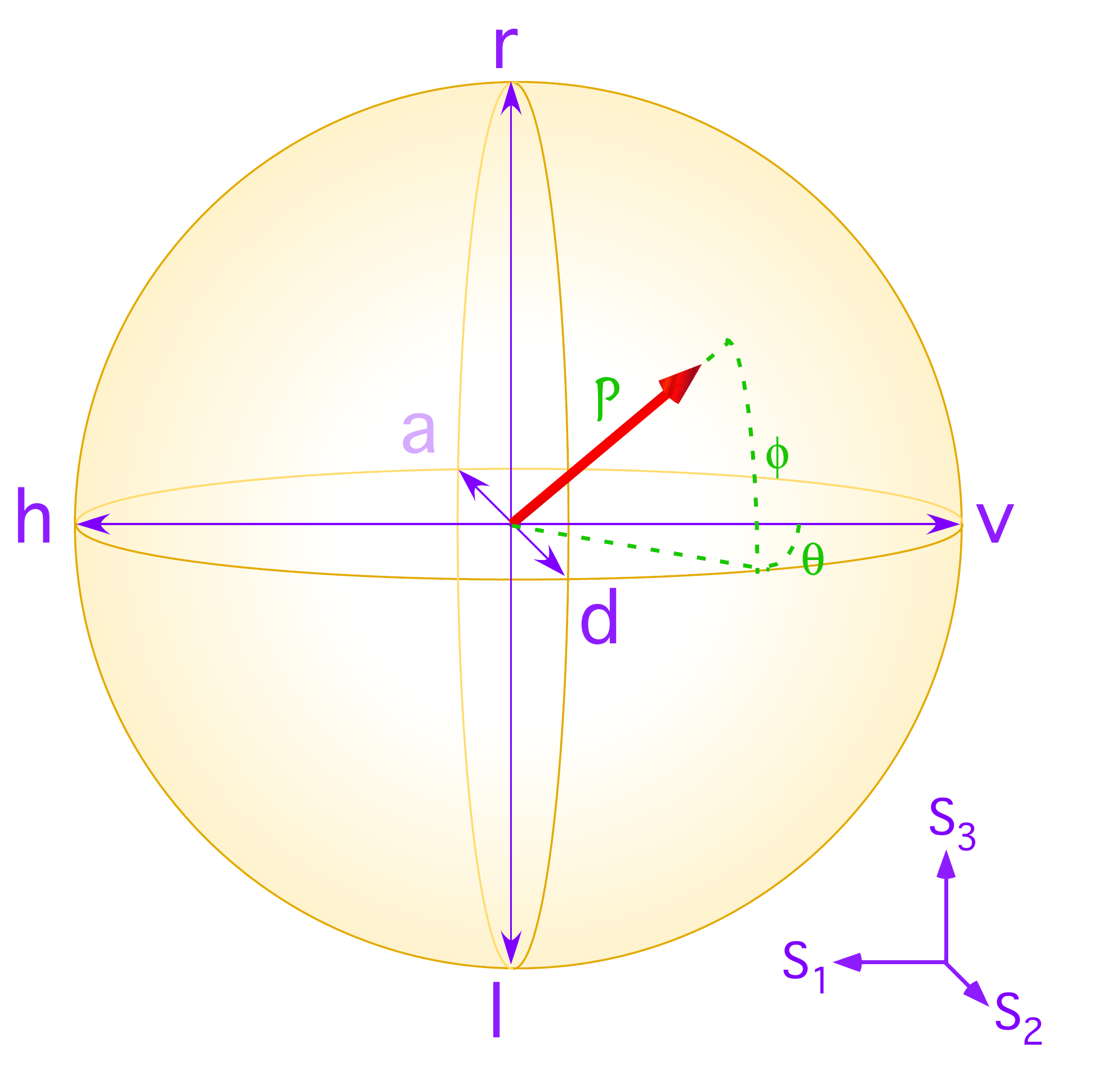}
\end{center}
\caption{Poincar\'e sphere. Any polarisation state of light can be represented by a Stokes' vector, \emph{red}, lying on (fully-polarised) or in  (partially-polarised) the sphere. The cartesian co-ordinates of the vector are given by Stokes' parameters \cite{Stokes}, $\{ S_{1}, S_{2}, S_{3} \}$, \emph{purple}; the end points of the axes are the horizontal/vertical ($h,v$), diagonal/anti-diagonal ($d,a$), and right/left-circular ($r,l$), polarised states, respectively. Alternatively the vector can be represented in spherical co-ordinates by a length, $\mathcal{P}$, and two angles, $\theta$, $\varphi$, \emph{green}. The vector length is the degree of polarisation, $\mathcal{P}{=}\sqrt{S_{1}^{2}{+}S_{2}^{2}{+}S_{3}^{2}}$; $\theta$ is the longitude, and $\varphi$ is the latitude.  For linearly polarised light, $\varphi{=}0^{\circ}$; for circularly polarised light. $\varphi{=}90^{\circ}$.}
\label{sphere}
\end{figure*}
\vfill

\newpage
\begin{figure*}
\begin{center}
\includegraphics[width=0.49 \linewidth]{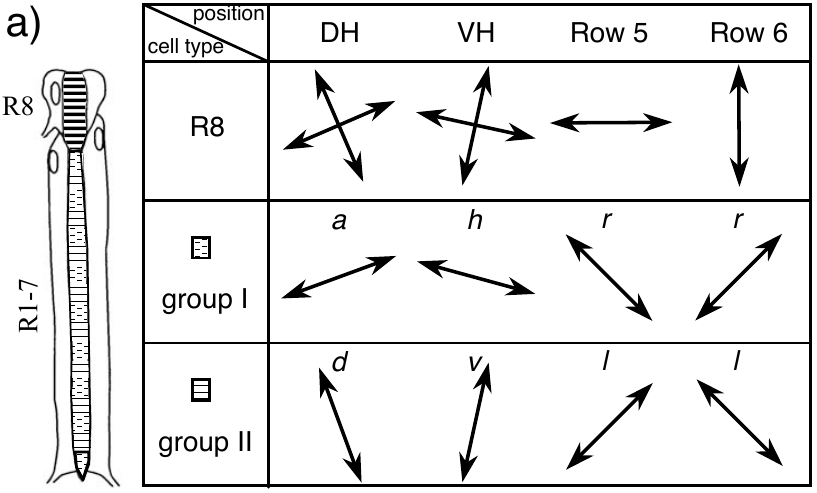}
\includegraphics[width=0.49 \linewidth]{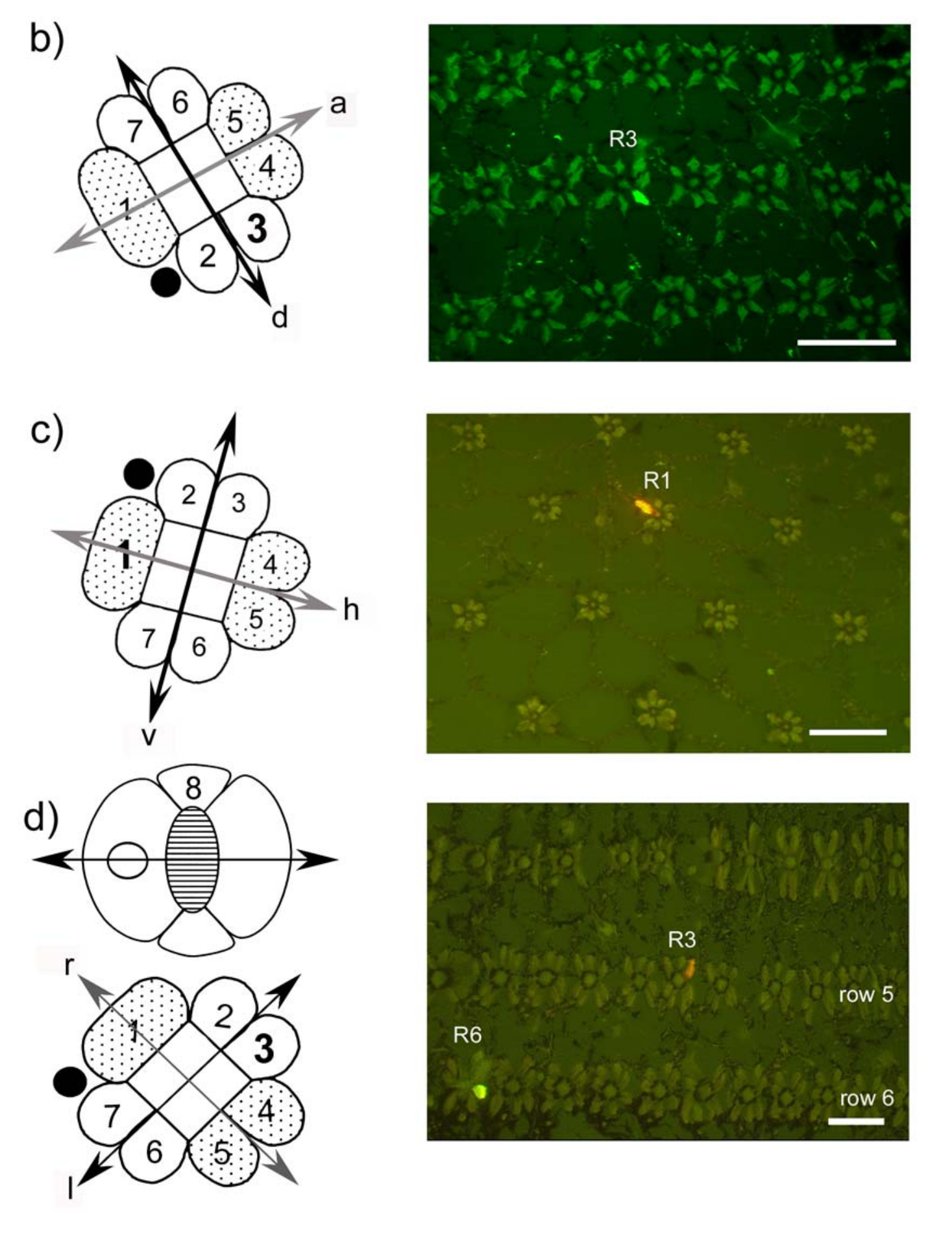}
\end{center}
\caption{a) \emph{left}. Diagram of a longitudinal section through an ommatidium (visual unit) in the hemispheres and mid-band rows 5 and 6 of the eye. 
The main rhabdom is formed by seven photoreceptors (R1--7), overlayed by a small, four-lobed ultraviolet sensitive photoreceptor, R8. \emph{right}. The arrows indicate the microvillar directions within each retinal region as if looking into the eye (frontal view). The R1--7 cells are divided into group I cells (R1, R4, R5) and group II cells (R2, R3, R6, R7), which form layers of orthogonal microvilli throughout the rhabdom, Fig \ref{eye}c, and thus are sensitive to orthogonal polarisations. The overlying R8 cells in rows 5 and 6 are extraordinarily long and they produce parallel microvilli, whereas the R8 cells in the remainder of the retina produce microvilli that are both orthogonal and interdigitating (crossed arrows). $\{ r, l \}$ indicates sensitivity to right- and left- circular polarised light, $\{ a, d, h, v \}$ indicates sensitivity to anti-diagonal, diagonal, horizontal and vertical linear polarised light. All directions indicated in the text and subsequent figures refer to a frontal view of a right eye with the mid-band arranged horizontally. b)-d) Frontal diagrams of the R1--7 (numbered 1--7) cell body arrangement around the central light guide, \emph{rhabdom}; and examples of dye-injected cells shown in the photomicrographs (scale bars 50 $\mu$m). b) The dorsal hemisphere, which analyses $d,a$; c) the ventral hemispheres, which analyses $h,v$; and d), mid-band row 5 which analyses $r,l$. Group I retinular cells are stippled and group II retinular cells are plain. Bold numbers in the diagrams indicate the stained cell(s) in the accompanying photograph. Arrows indicate microvillar axes, and thus the directions of linearly polarised light to which the photoreceptors respond maximally, $\phi_{\mathrm{max}}$. Grey and black arrows indicate group I and II receptors, respectively. The cell arrangement in mid-band row 6 (not shown) is rotated 90$^{\circ}$ counter-clockwise compared to row 5. d) Circular polarisation sensitivity is not innate to the R1--7 cells, but arises from the quarter-wave retardance of the overlying four-lobed R8 cell, \emph{top}, effective for $400{<} \lambda {<} 700$ nm. Quarter-wave retardance is realised by increased photoreceptor length and by the formation of unidirectional microvilli, the axis of which is indicated by the black arrow. The R8 microvilli are arranged at 45$^{\circ}$ to the underlying orthogonal microvillar sets formed by the R1--7 cells, \emph{bottom}: R8 converts circular polarised to linear polarised light at ${\pm}45^{\circ}$ to the R8 microvillar axis, depending on the handedness of the circular polarisation. 
Both dye-filled photoreceptors (row 5 R3 and row 6 R6) belong to group II receptors. The angle between the microvillar (optical) axes of the R8 cells and the microvillar directions of group II photoreceptors is ${-}45^{\circ}$ in both mid-band rows 5 and 6. Both stained retinular cells are therefore more sensitive to $l$; similarly the group I photoreceptors are more sensitive to $r$.
}
\label{rhabdom}
\end{figure*}



\newpage
\begin{figure*}
\begin{center}
\includegraphics[width=\linewidth]{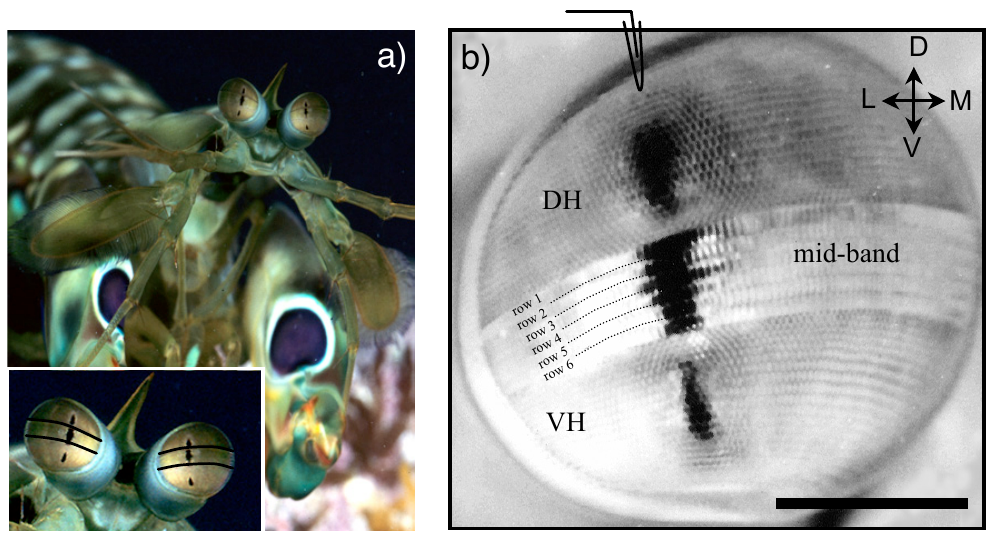}
\includegraphics[width=\linewidth]{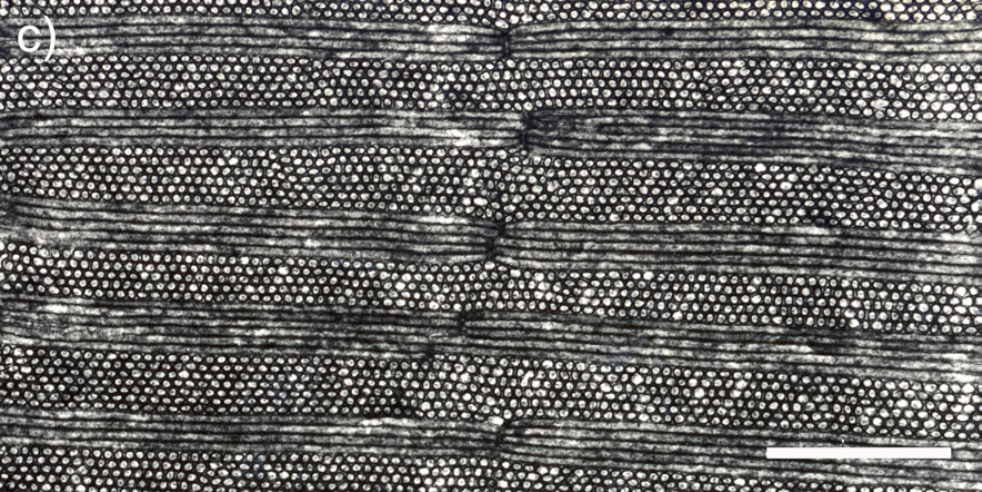}
\end{center}
\caption{a) Adult \emph{Gonodactylus smithii}, or mantis shrimp, ${\sim}$7 cm long. The stalked apposition compound eyes are divided into a dorsal and a ventral hemisphere by an equatorial mid-band of enlarged and structurally specialised ommatidia. \emph{Inset}. Mid-band position indicated by curved dark lines. The three pseudopupils (dark spots) visible within each eye indicate that the visual fields of the two hemispheres and the mid-band almost completely overlap at the equator of the eye, so that the three eye regions view the equatorial strip simultaneously. \emph{Photograph by R.L. Caldwell.} b) Frontal view of the right eye to illustrate the division of the eye into a dorsal hemisphere (DH) and a ventral hemisphere (VH) by the equatorial mid-band formed by six rows of enlarged ommatidia, numbered row 1 to row 6 from dorsal to ventral. Mid-band rows 1-4 contain spectral photoreceptors; mid-band rows 5 and 6 are specialised for circular polarisation vision; the dorsal and ventral hemispheres for linear polarisation, as described in Fig.~\ref{rhabdom}a. Recording electrodes were lowered through corneal holes cut in the lateral half of the dorsal hemisphere, where the mid-band is ${\sim}15^{\circ}$ relative to the equator of the eye. The black scale bar is 1 mm, the axes refer to Dorsal, Medial, Ventral and Lateral. c) Electron micrograph of a longitudinal section through a mid-band row 6 rhabdom. The alternating layers of microvilli are highly ordered and in thinner layers than in hemispheric rhabdoms. Mid-band rows 5 \& 6 retinular cells have twice the degree of polarisation of hemispheric cells due to a more crystalline microvillar structure: c.f. $\overline{\mathcal{D}}_{\mathrm{mid}}{=}0.340{\pm}0.061$ with $\overline{\mathcal{D}}_{\mathrm{hemi}}{=}0.145{\pm}0.035$. The white scale bar indicates 1$\mu$m.
}
\label{eye}
\end{figure*}
\vfill

\newpage
\begin{figure*}
\begin{center}
\includegraphics[width=0.4 \linewidth]{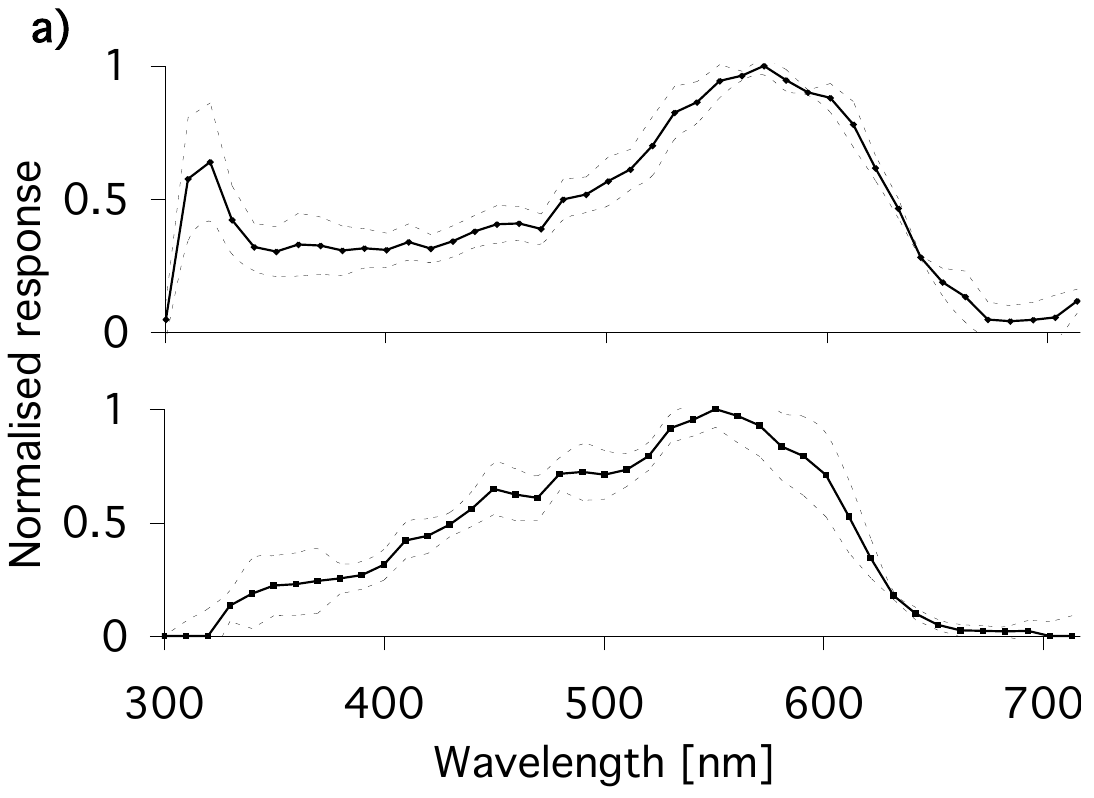} \hspace{10 mm}
\includegraphics[width=0.49 \linewidth]{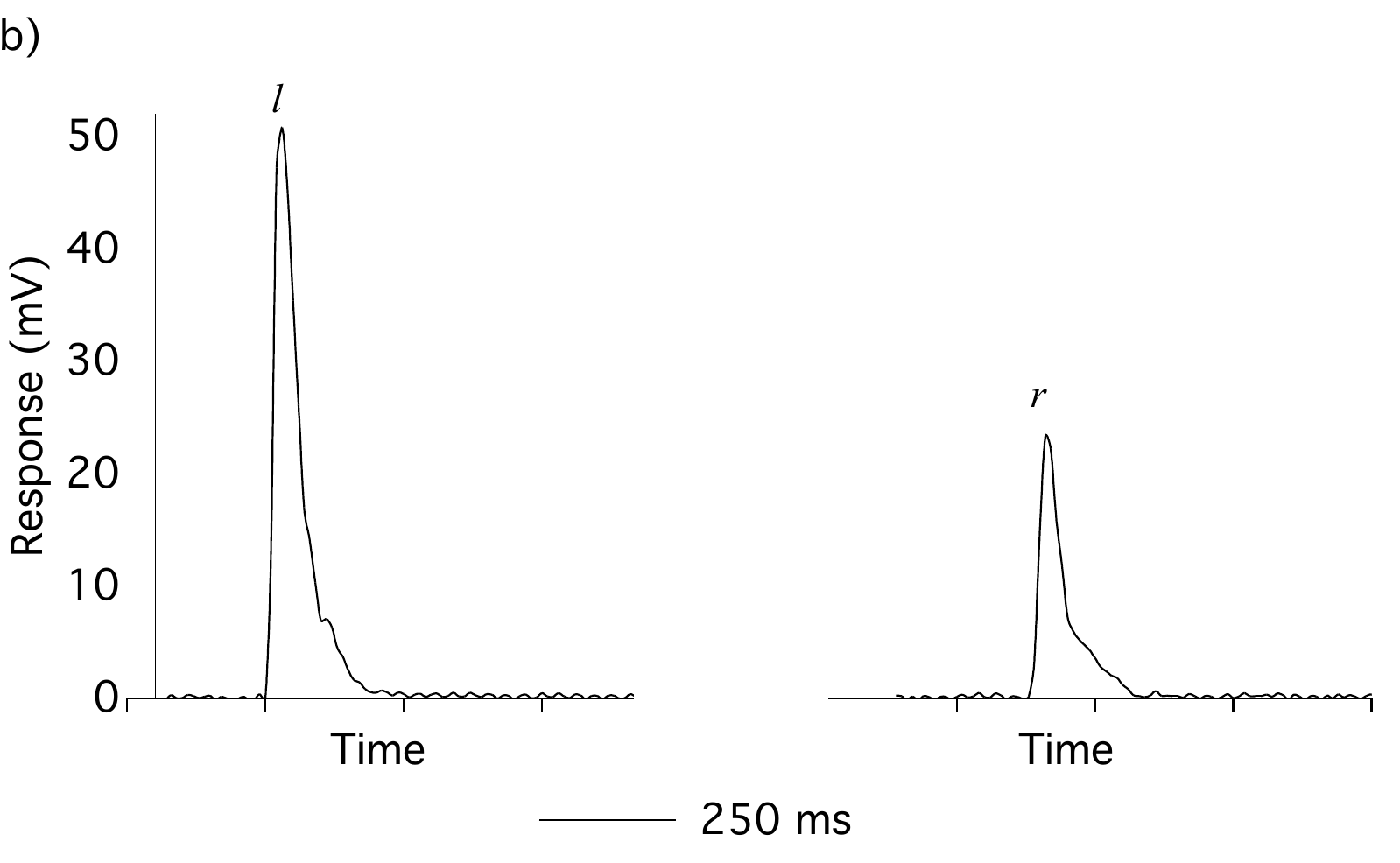}
\includegraphics[width=0.49 \linewidth]{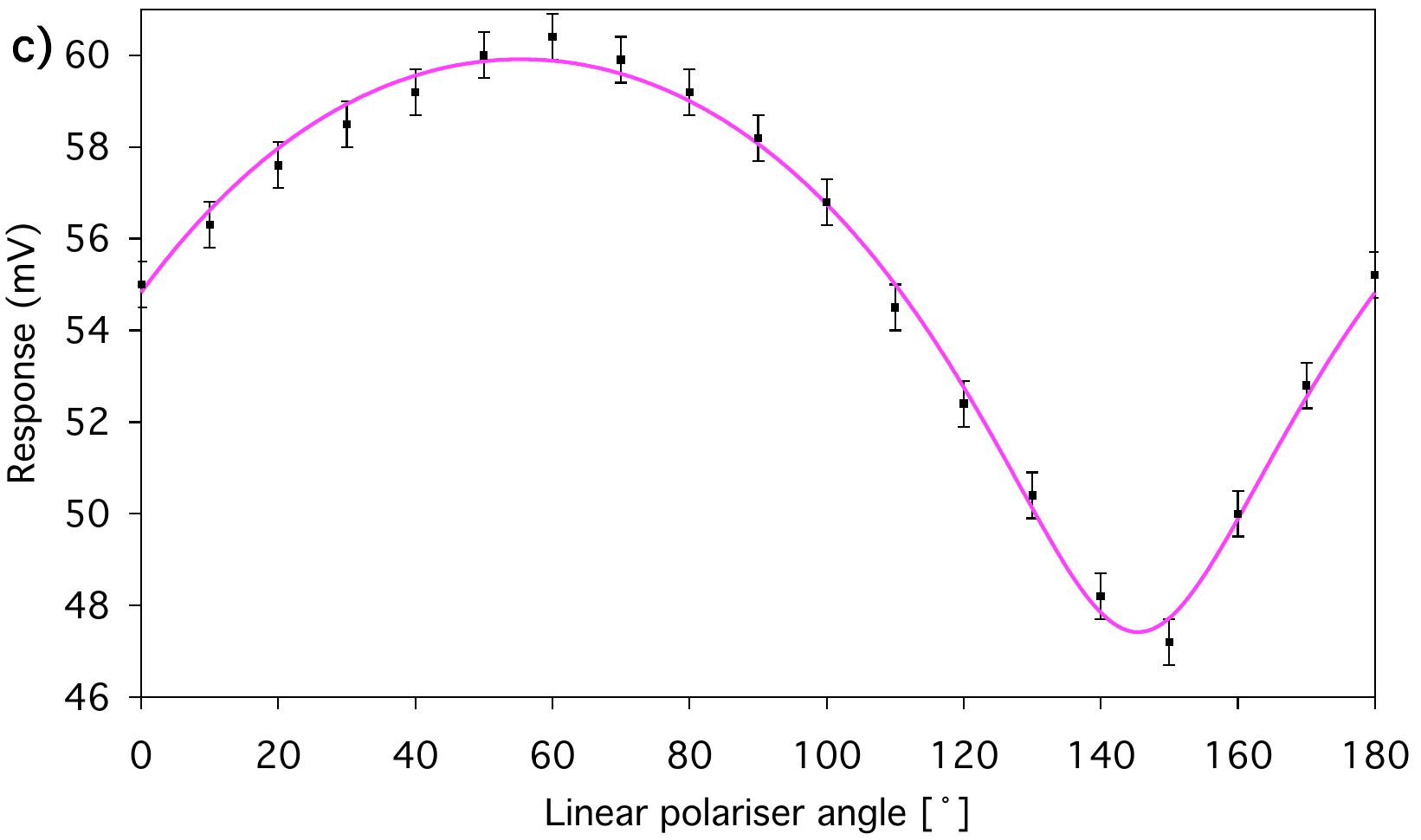}
\includegraphics[width=0.49 \linewidth]{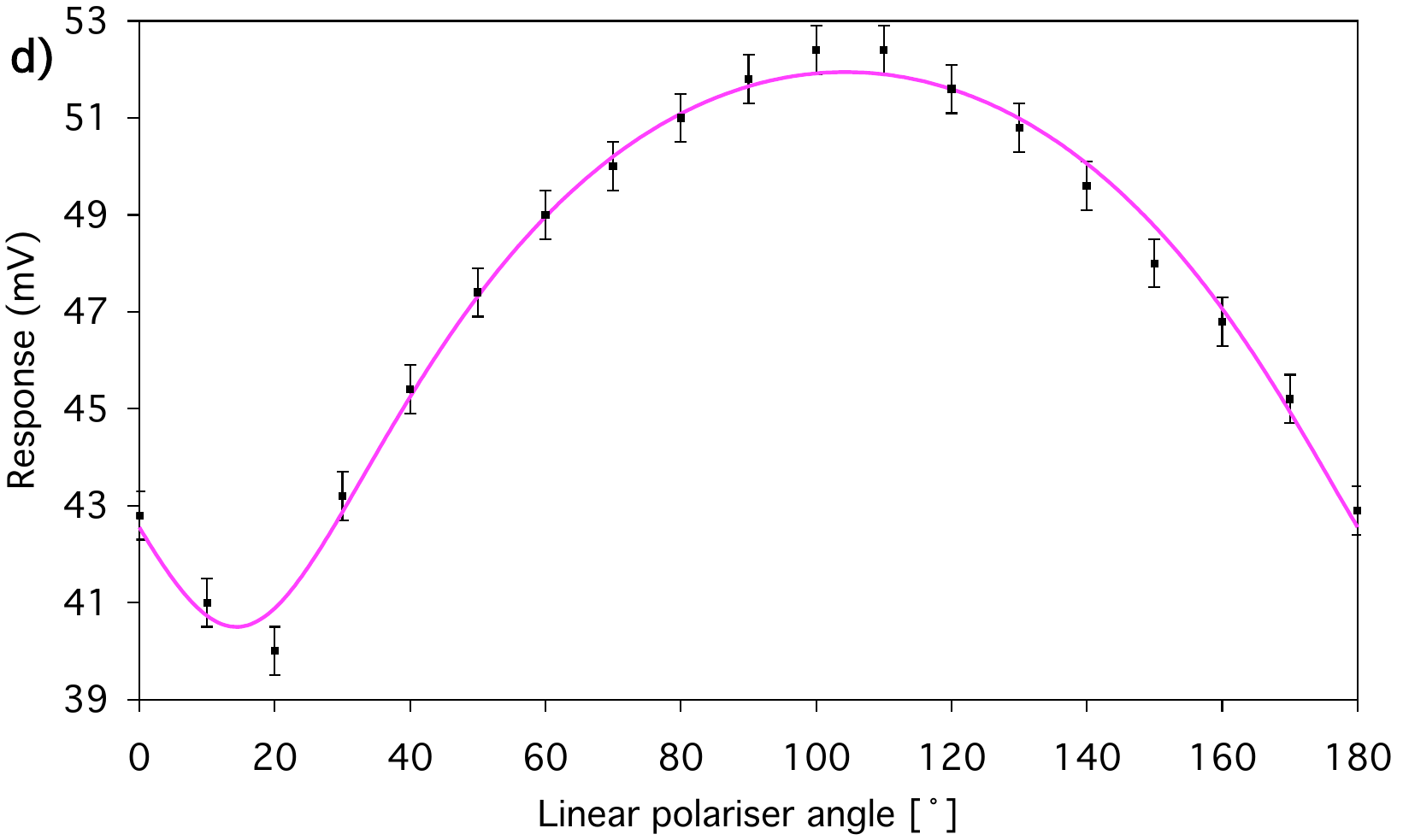}
\includegraphics[width=0.49 \linewidth]{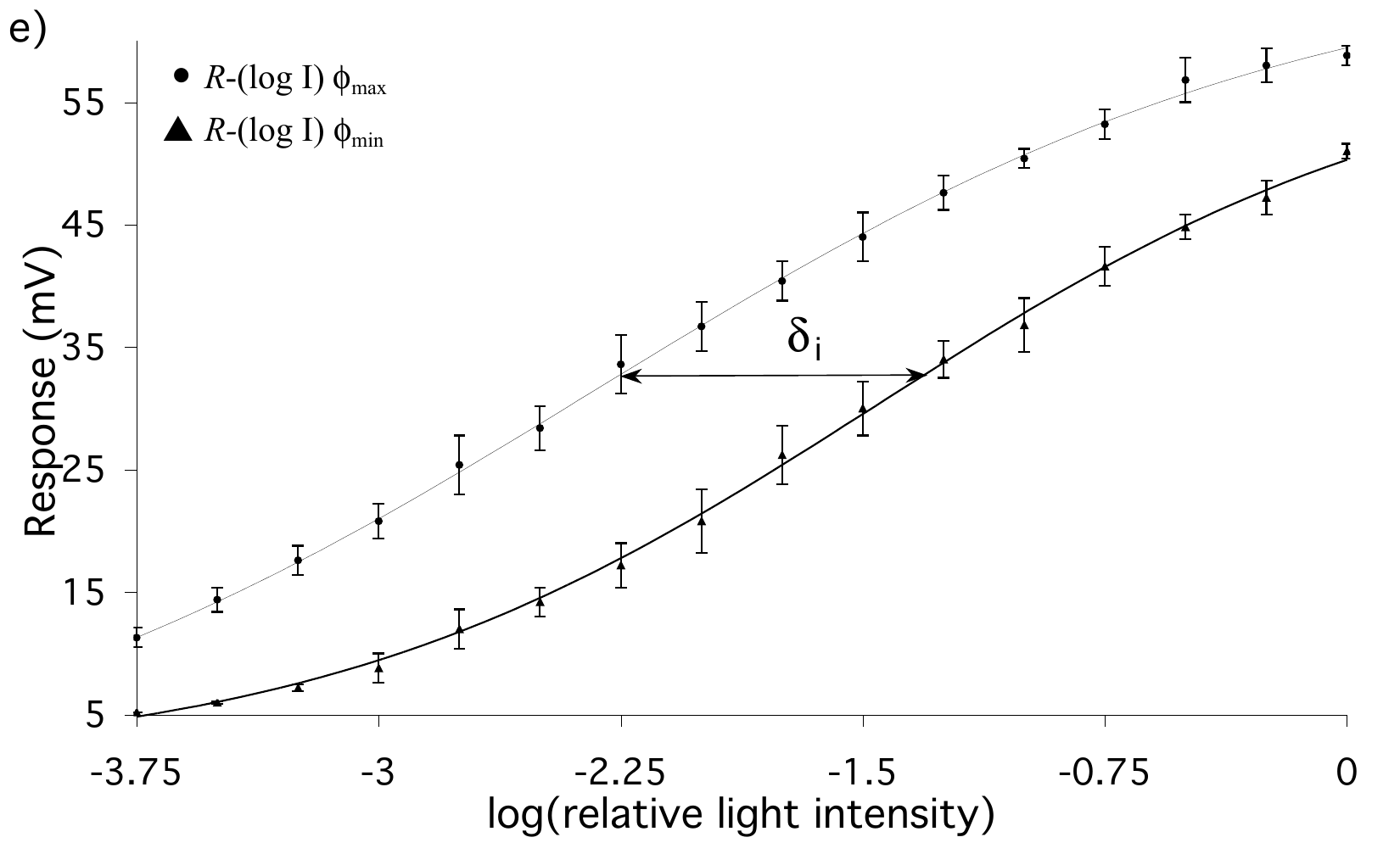}
\includegraphics[width=0.49 \linewidth]{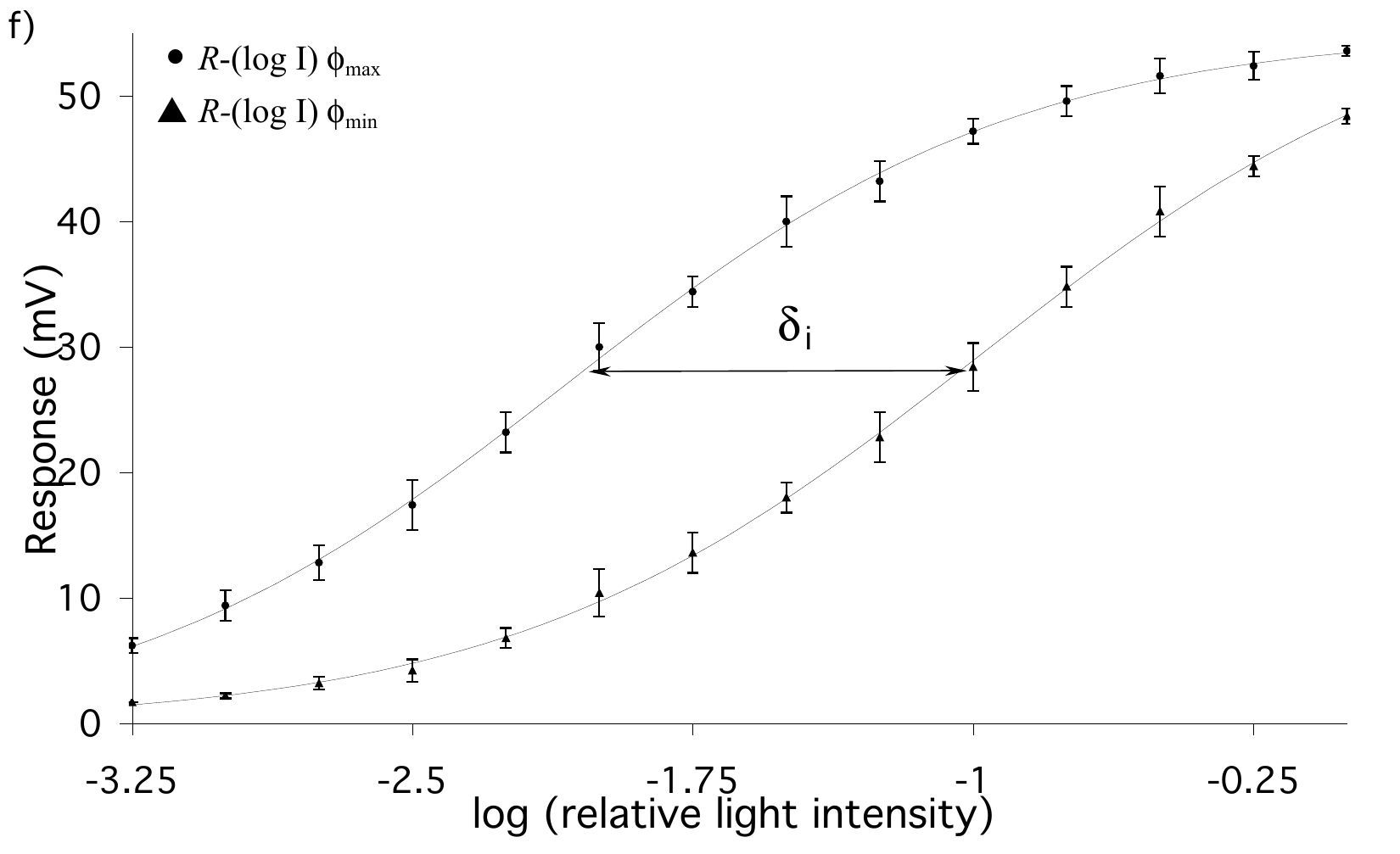}
\includegraphics[width=0.49 \linewidth]{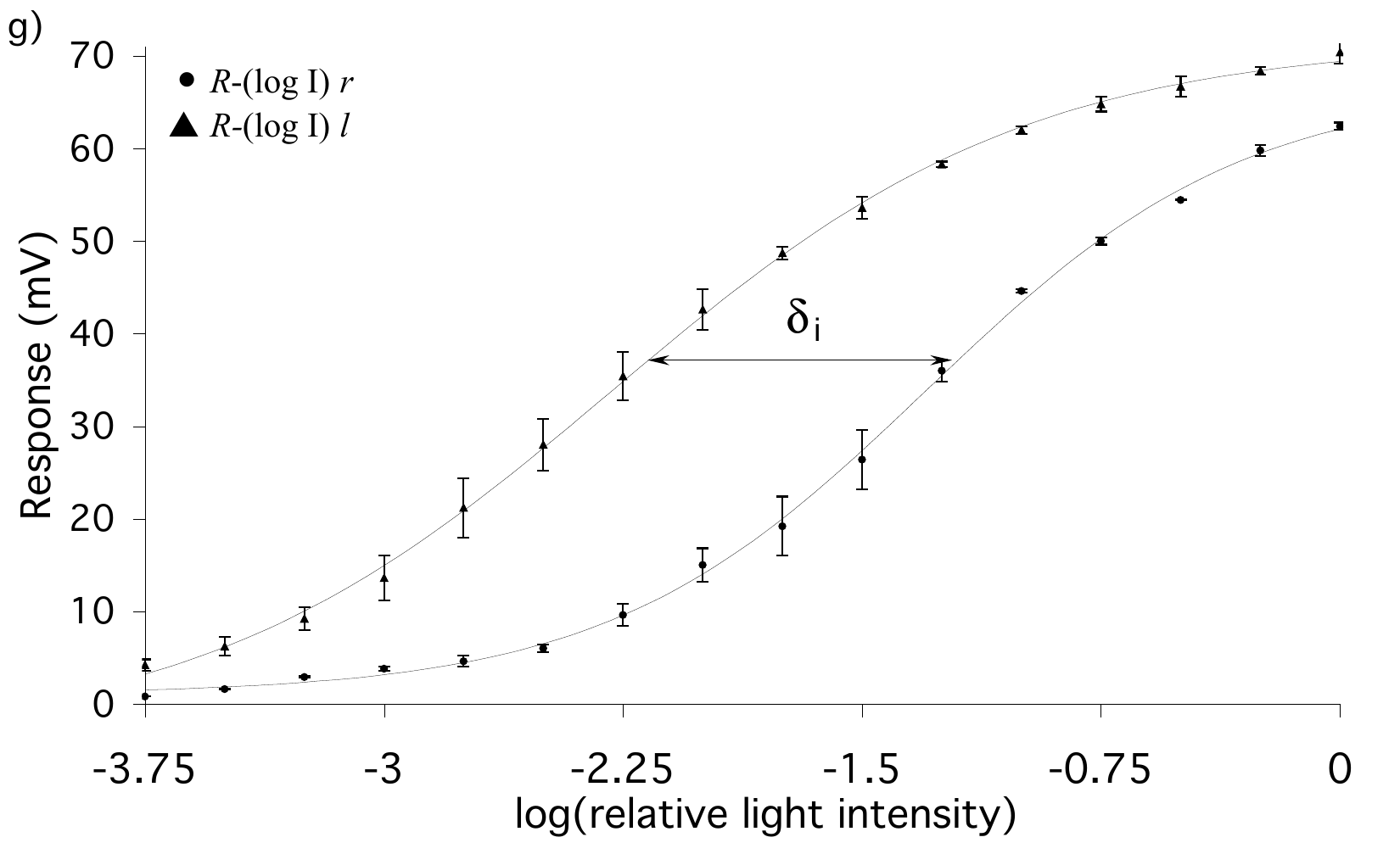}
\includegraphics[width=0.49 \linewidth]{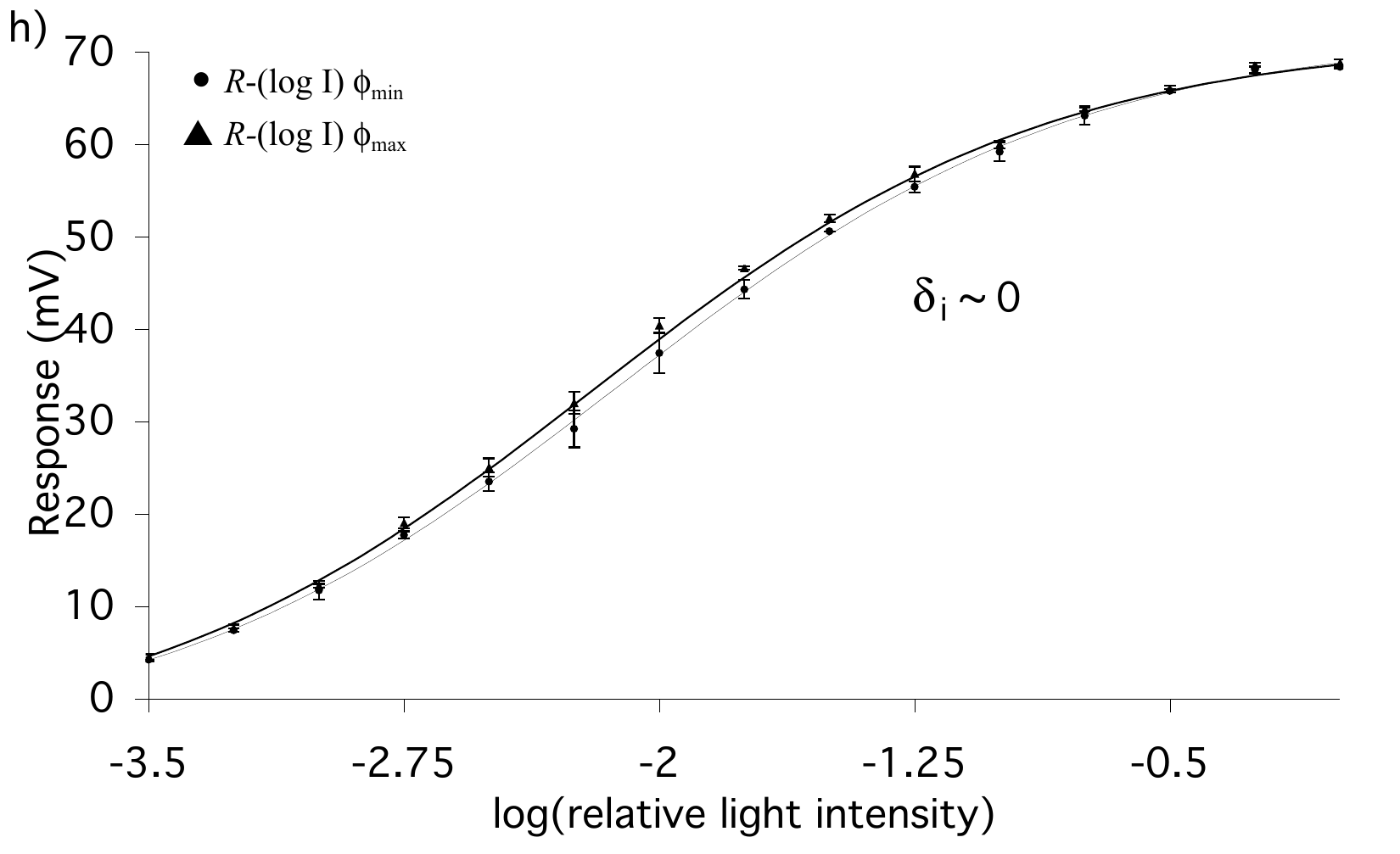}
\end{center}
\caption{a) Normalised spectral responses averaged over: \emph{top}) 12 hemispheric cells and \emph{bottom}) 14 mid-band cells, rows 5 \& 6. b) Response vs time for a row 5 R3 photoreceptor when illuminated with a 50 ms spectrally-filtered light pulse: \emph{left}) left-circular, $l$, and \emph{right}) right-circular, $r$, polarisation. This cell responds more strongly to $l$. c), d), Peak response versus polariser angle for R1 cells in the c) dorsal and d) ventral hemispheres. These act as anti-diagonal and horizontal polarising photoreceptors, respectively. The smooth pink lines are logarithmic cosine-squared curves with 4 fit parameters (phase, cosine amplitude, response offset, and logarithmic amplitude). e), f), Peak response versus relative light intensity for the c) and d) cells. Top curves are measured at the maximum response angle, $\phi_{\mathrm{max}}$;  bottom at $\phi_{\mathrm{min}}$. The smooth black lines are sigmoidal curves fitted to a Naka-Rushton function using least-squares approximation in Origin 6.1. The polarisation sensitivity, $10^{\delta_{i}}$, is measured by taking the difference between intensities in the linear part of the curves, $\delta_{i}$. g), h), Peak response versus relative light intensity for a row 5 R1 cell. g) Response to left and right circular polarised light. h) Response to linearly polarised light ${\pm}45^{\circ}$ from the vertical.}
\label{data}
\end{figure*}
\vfill

\begin{figure*}
\begin{center}
\includegraphics[width=0.5 \linewidth]{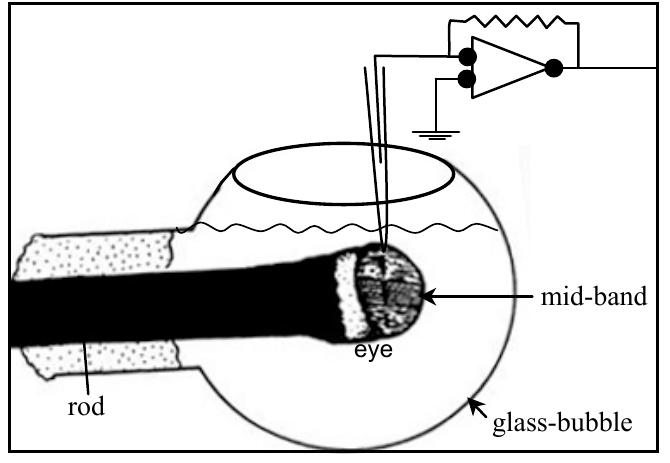}
\caption{For electrophysiological recordings the isolated eye was mounted on a plastic rod and placed into a glass bubble filled with stomatopod saline so that the lateral mid-band was oriented horizontally. The intracellular electrode was lowered vertically through a small hole cut into the lateral cornea of the dorsal hemisphere.}
\label{eyemount}
\end{center}
\end{figure*}

\begin{figure*}
\begin{center}
\includegraphics[width=0.5 \linewidth]{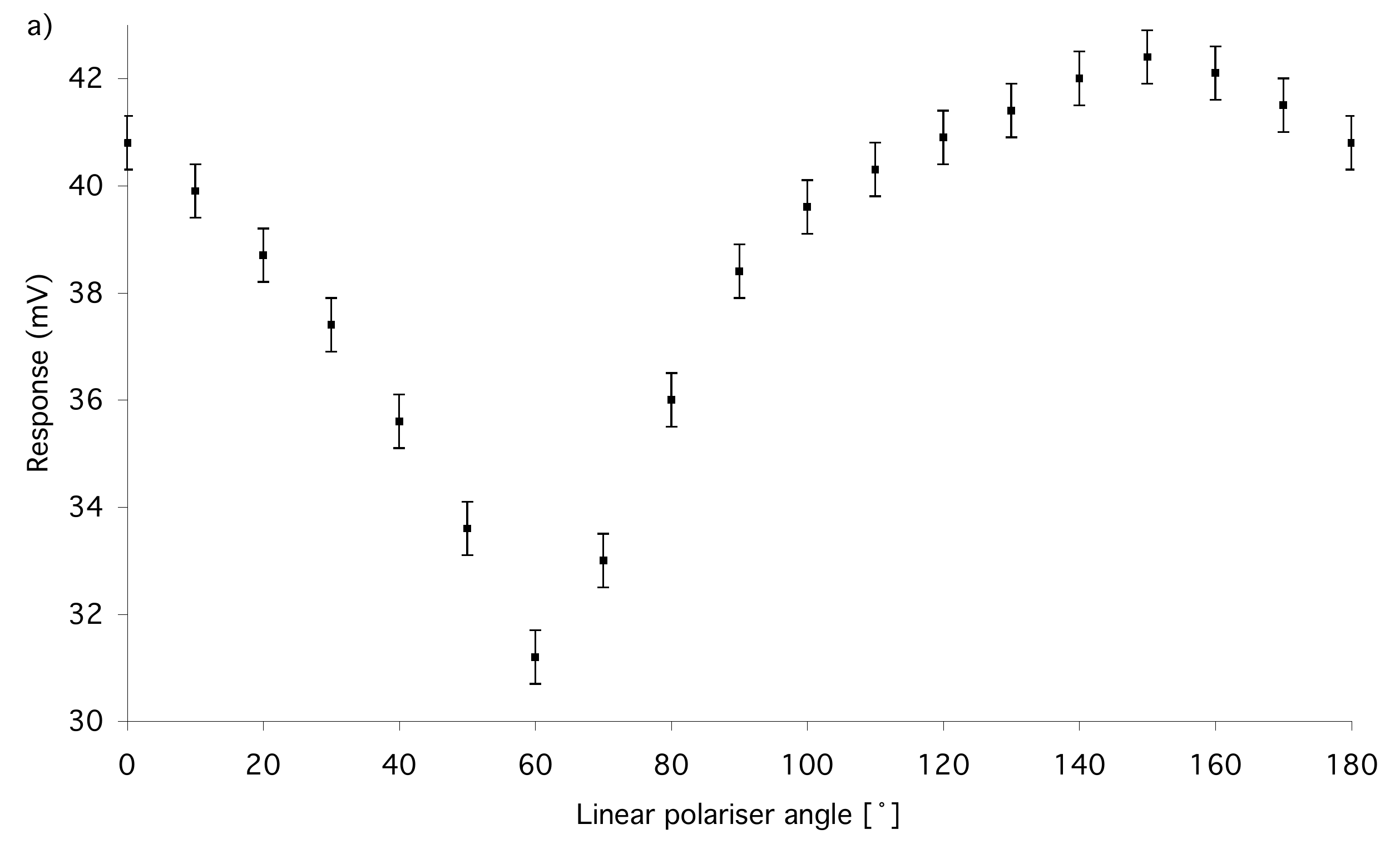}
\includegraphics[width=0.5 \linewidth]{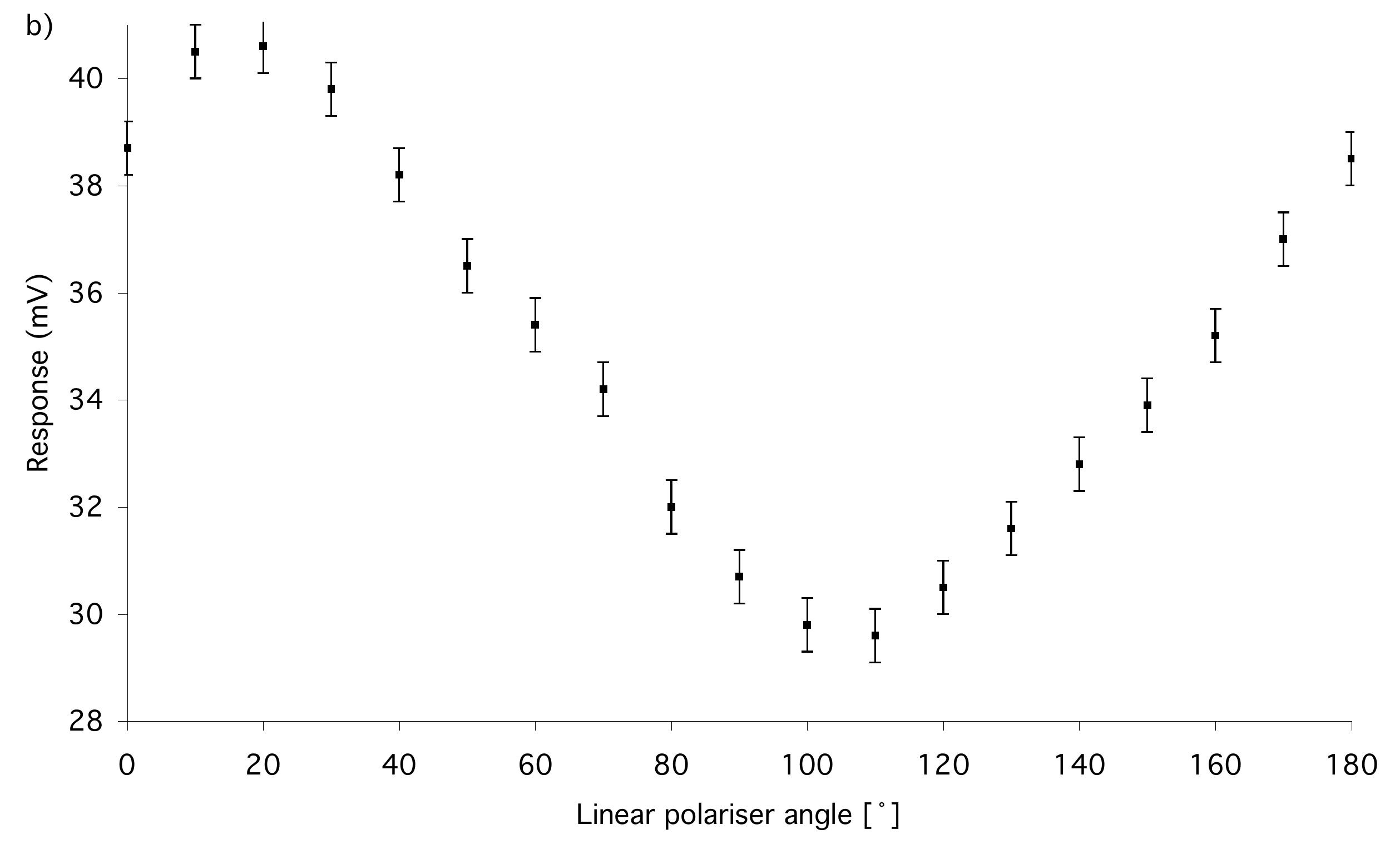}
\caption{Peak response versus polariser angle for a) the R3 cell in the dorsal and d) the R7 cell in the ventral hemispheres, respectively. These group II cells act as diagonal and vertical polarising photoreceptors, respectively, sensitive to the orthogonal polarisations to the group I cells shown in Figs 4c),d) in the main  text.}
\label{curves}
\end{center}
\end{figure*}

\end{document}